\documentstyle[emulateapj]{article}
\input{psfig}

\lefthead{Escala}
\righthead{On the Fueling of Massive Black Holes and the Properties of their Host Spheroids}

\def\msun{M_{\odot}}

\begin{document}

\title{On the Fueling of Massive Black Holes and the Properties of their Host Spheroids} 
\author{Andr\'es Escala}
\affil{Departamento de Astronom\'{\i}a, Universidad de Chile, Casilla 36-D, Santiago, Chile.}

\begin{abstract}
We study the relation between nuclear massive black holes and their host spheroid gravitational potential. Using simple  models, we analyze how gas is expected to be transported in the nuclear regions of galaxies. When we  couple it with the expected gas lifetime given by the  Kennicutt-Schmidt Law, this naturally leads to the `$\rm M_{BH}$ - $\rm M_{virial}$'  and `$\rm M_{BH}$ - $\sigma$' relations. 
\end{abstract}


\section{Introduction}

Most nearby massive spheroids (elliptical and spiral bulges) host nuclear massive black holes (MBH), whose  mass scales as $\rm M_{BH} \propto \sigma^{4-5}$ (the `$\rm M_{BH}$ - $\sigma$' relation; Ferrarese \& Merritt 2000;  Gebhardt et al. 2000). In recent years, several theories have been proposed  to clarify the origin of this relation (Silk \& Rees 1998; Blandford 1999; Fabian 1999; Burkert \& Silk 2001; Zhao, Haehnelt \& Rees 2002; King 2003; Adams et al. 2003; Miralda-Escaud\'e \& Kollmeier 2005; Sazonov et al. 2005; Begelman \& Nath 2005). However, very little effort has been devoted to study how angular momentum is transfered in the centers of galaxies. This is an unavoidable challenge for the understanding of the formation of MBHs and the origin of the `$\rm M_{BH}$ - $\sigma$' relation. 

The study of MBH growth by gas accretion is usually  focused  on the study of accretion disks. However, these accretion disks are Keplerian by nature and therefore  have neglible masses compared to that if the   MBH. They must be continously replenished, otherwise the mass  of the MBH  will not  have a considerable growth. The key question in the growth of MBHs by accretion, is  how to remove the large angular momentum of gas in a galaxy in order to funnel it into the accretion disk in the central sub-pc region: the so called  `Fueling Problem'. Without a proper  understanding of this `Fueling Problem', specially in the last few 100 pc, it  is impossible to understand the growth of MBHs and to elucidate why their masses correlate with the host spheroid properties. 


\section{Mass Transport in Galactic Nucleus}

Several mechanisms for fueling gas from R $\sim$10 kpc  down to 100 pc have been proposed in the literature, including galaxy interactions, stellar bars and gravitational resonances (see Shlosman, Begelman \& Frank 1990 for a classic review, or Wada 2004 for a more recent one). However, gravitational torques in galaxy mergers arise as  the dominant process for fueling large amounts of gas down to the central few hundred parsecs and for triggering most of the MBH growth.  In a merger, after a violently relaxed core is formed at the center, most of the gas  will settle  in a nuclear disk (with typically several hundred parsecs in diameter) 
  that is  rotationally supported against the overall 
gravitational potential (Barnes 2002). The MBHs will migrate to the center, and merge in a timescale relatively short compared to the lifetime of the nuclear disk (Escala et al. 2004, 2005).

\subsection{Turbulent Disk}

Nuclear gas disks are characterized by being self-gravitating. It has been debated whether gravitational instability leads to disk fragmentation (`disk of clouds'; Shlosman et al. 1990) or to transient density fluctuations (`turbulent disk';  Wada 2004),  depending on cooling/heating rates and thermal physics.
 I will mainly focus in the latter case because it is possible to advance further with analytical estimates. 

A turbulence-dominated disk  can be  modeled on a zero-order approximation as a steady $\alpha$-disk. We choose  to parameterize  the viscosity $\rm \nu$   in terms of the turbulent speed  $\rm  v_{turb}$ instead of the  sound speed, because the turbulence in a nuclear disk is expected to be highly supersonic (Downes \& Solomon 1998) and therefore  $\rm  v_{turb}$  becomes  dynamically more relevant. In such a case, the disk height  {\it h} is $\rm v_{turb}/\Omega$ (being $\rm \Omega$  the angular velocity) and the viscosity $\rm \nu$,  parameterized by Shakura and Sunyaev as $\rm\alpha\Omega {\it h}^2$, becomes $  \nu =  \alpha v_{turb}^2/\Omega$. Therefore in a turbulence-dominated $\alpha$-disk with a flat rotation curve, the mass accretion rate can be written as 
\begin{equation}
\rm \dot{M} = \frac{2\sqrt{2}\, \alpha \, v_{turb}^{3}}{G\, Q} \, ,
\label{mdot1}
\end{equation}
where  $\rm Q \equiv v_{turb} \Omega / \pi G \Sigma_{g}$  is the {\it `turbulent'} Toomre's parameter and $\rm \Sigma_{g}$ is the gas surface density. 
 Eq. ($\ref{mdot1}$) represents the mass accretion rate  onto the inner accretion disk around the MBH (fueling).

In nuclear disks the turbulence can be driven by two mechanisms: feedback from star formation and self-gravity. The thickness of these nuclear disks  is independent of whether or not the disk hosts a starburst (Jogee et al. 2005),  suggesting that the turbulence is  driven by self-gravity. 
 We will now focus on the  case where the turbulence is driven by gravitational instabilities and the energy is extracted from the disk rotation. Unfortunately, there is no consensus on the energy spectrum $\rm E(k) \propto k^{n}$ for compressible supersonic turbulence (Ballesteros-Paredes et. al 2006). Boldyrev (2002) has presented a theoretical model suggesting that $\rm n \approx -1.74$ and that appears to be confirmed numerically (Joung \& Mac Low 2006 and references therein). Boldyrev's theory suggests that in the inertial range, compressible supersonic turbulence obey the Kolmogorov Law. 
In such a case  the energy is transported in a cascade from the largest unstable scale 
characterized by the disk radius $\rm R_{d}$ and rotation velocity $\rm  v_{rot}$, to the turbulent eddies' scale  characterized by the size $\rm \lambda_{turb}$ and velocity $\rm  v_{turb}$. This  can be expressed as 
\begin{equation}
\rm  v_{turb} \sim v_{rot} \left( \frac{\lambda_{turb}}{R_{d}} \right)^{1/3} \, .
\label{kolgov}
\end{equation}

We shall also consider that the gas is in a disk in  rotational support against the overall gravitational potential. For simplicity we  assume  that   $\rm  v_{rot} \, and \, \sigma$ are constant with radius, that is  only strictly valid for an isothermal sphere. 
On the central kiloparsec of galaxies, stellar bulges are   the dominant components of the  gravitational potential even in  the most massive gaseous nuclear disks known so far, where observations (Downes \& Solomon 1998) found typically 5 times more mass in the bulge that in the gas ($\rm M_{gas}/M_{star} \sim 1/5$). Even in such an extreme case, the rotation velocity is  $\rm  v_{rot} = (  1.2 \,  G \, M_{star}/R)^{1/2}  = 1.09 \, \sigma \approx \sigma$  


Finally,  combining  $\rm  v_{rot} \approx \sigma$ with Eqs. $\rm (\ref{mdot1})\, and \, (\ref{kolgov})$  we find that the  mass accretion rate onto the inner Keplerian accretion disk is
\begin{equation}
\rm \dot{M} \sim \frac{2\sqrt{2}\, \alpha}{G\, Q}  \left( \frac{\lambda_{turb}}{R_{d}} \right) \, \sigma^{3}   \, .
\label{mdot2}
\end{equation}

Equation $\rm (\ref{mdot2})$ shows that the  mass transport  in the nuclear regions of galaxies is controlled by the spheroid velocity dispersion $\sigma$ (the other factors are either dimensionless or physical constants). Therefore the rate at which the inner  accretion disk around the MBH is  replenished, strongly depends on the spheroid that hosts the MBH. 



\subsection{Disk of Clouds}

A disk composed by randomly moving clouds  is the second possible model for the nuclear disk. The accretion flow in this `disk', can be expressed in the same form of Eq. $\ref{mdot1}$    replacing  $\rm v_{turb}$   by the cloud velocity dispersion $\rm \Delta v$ (Goldreich \& Tremaine 1978). The formation of the clouds is driven by gravitational instabilities and their orbital energy is extracted from the disk rotation, as in the gravity-driven turbulence case. Therefore we expect that $\rm \Delta v$ scales with $\rm v_{rot}$ in a similar way as $\rm v_{turb}$ does, leading to an expression for the mass accretion rate analogous to Eq. $\ref{mdot2}$. However, the exact value of $\rm \Delta v$ is highly uncertain and numerical simulations are required for an accurate estimation.   

\section{Closure Equation: Gas Lifetime}

In order to determine the total mass accreted, we need to know for how long gas will be  available  to be accreted by the MBH.

Usually, AGN feedback is considered as the most likely physical process that determines the characteristic   gas lifetime, expelling most of the gas in the host galaxy when  the Eddington Limit is exceeded.  However, in an inhomogeneous medium this limit can be greatly exceeded because photons tend to diffuse trough the low density medium. Several instabilities will produce inhomogeneities large enough to guarantee that the Eddington Limit is no longer valid, such as photon-bubble (Begelman 2001) or Kelvin-Hemboltz instabilities (Krumholz et. al 2005). Moreover on galactic scales,  the ISM is endeed highly inhomogeneous and  is  hard to believe that most of the gas in a galaxy  will be expelled  due to  AGN feedback.     

On the other hand, these massive nuclear disks are characterized by an ongoing starburst and therefore gas depletion due to star formation will eventually be relevant in the absence of other processes. Although star formation is still poorly understood, it can be parameterized by a well constrained empirical relation called the Kennicutt-Schmidt Law (Kennicutt 1998)  which can be stated as
\begin{equation}
\rm \dot{\Sigma}_{gas} = 0.017 \,  \frac{\Sigma_{gas}}{\tau_{dyn}} \, ,
\label{kenn}
\end{equation}
where $\rm \dot{\Sigma}_{gas}$ is the star formation rate per unit area, $\rm \Sigma_{gas}$ is the gas surface density and $\rm \tau_{dyn} = R_{d}/v_{rot}$ is the dynamical (i.e. orbital) timescale of the gas. 
These  nuclear disks have typical radius of $\rm R_{d} \sim 0.1 \, R_{e}$, being $\rm R_{e}$ the  effective radius of the host spheroid. As shown in \S 2, the bulge dominates the overall gravitational potential ($\rm  v_{rot} \approx \sigma$) and therefore the dynamical timescale of the gas can be expressed as $\rm \tau_{dyn} = 0.1 \, R_{e}/\sigma$. Thus  Eq. $\ref{kenn}$ implies that the gas lifetime  $\rm t_{gas}$, defined as $\rm \Sigma_{gas}/\dot{\Sigma}_{gas}$, is   $\rm  t_{gas} = \xi \, R_{e}/\sigma \,\,$ where $\rm \xi \approx 5.9$. 

The final mass of the MBH will be determined by $\rm M_{BH} = \eta  \, \dot{M} \,  t_{gas}$, where   $\rm \eta$ is the fraction of gas in the inner accretion disk that finally ends inside the event horizon of the MBH. Using Eq. ($\ref{mdot2}$) the final MBH mass becomes 
\begin{equation}
\rm M_{BH}  \sim \frac{2\sqrt{2}\, \eta  \, \xi \, \alpha}{\, Q}  \left( \frac{\lambda_{turb}}{R_{d}} \right)  \frac{R_{e} \, \sigma^{2}}{G}  
 \, ,
\label{mbh}
\end{equation}
which reproduces  the correlation between $\rm M_{BH}$ and the bulge  mass  $\rm M_{virial} = k \, R_{e} \, \sigma^{2}/G$ (Marconi \& Hunt 2003; Haring \& Rix 2004), being k=8/3 for an isothermal sphere. 

Using  the Faber-Jackson relation $\rm R_{e} \propto \sigma_{e}^{2}$ in Eq. $\ref{mbh}$, we obtain  the standard `$\rm M_{BH}$ - $\sigma$'  form ($\rm M_{BH} \propto \sigma^{4}$). Since the `$\rm M_{BH}$ - $\sigma$' and `$\rm M_{BH}$ - bulge virial mass' correlations have similar levels of scatter,  the Faber-Jackson relation should not  have larger scatter than either one in  the galaxies where the   correlations are measured. 


Finally, we need to evaluate the zero point in Eq. $\ref{mbh}$. For gravity-driven turbulence $\alpha$ may be as large as $\rm  0.3$ (Gammie 2001). In nuclear disks $\rm  v_{turb}/v_{rot} \sim 1/5$ (Downes \& Solomon 1998), implying that $\rm \lambda_{turb}/R_{d} \sim 0.008$ (Eq. $\ref{kolgov}$). Using a typical value of Q = 1.5 (Martin \& Kennicutt 2001), we find $\rm M_{BH} \approx 0.009 \, \eta \, M_{virial}$. To reproduce the `$\rm M_{BH}$ - bulge virial mass' relation, we need to have $ \eta \approx 1/6$. Thus, we  need that  less than 20\% of the mass fed onto the inner accretion disk ends up inside the event horizon of the MBH. 


\section{Numerical Study}

The conclusions found in \S 2 and \S3  are based in a steady-state model ($\alpha$-disk). However, previous numerical studies on the mass transport in the inner 100 pc of galaxies (Wada \& Norman 2002; Wada 2004) found that the mass accretion rate is highly time dependent with fluctuations of several order of magnitudes. For that reason, we have designed a  simple numerical  experiment to test the validity of the  $\alpha$-disk model as a zero-order approximation. 

In massive nuclear disks, self-gravity, cooling and feedback effects from star
formation play a major role in determining the physics of the ISM and therefore
 must be included to accurately model the angular momentum and mass transport
in the disk. However, a fully realistic model has multiple relevant parameters to be studied (such as the ones in  the star formation prescription, heating/cooling processes, etc) and requires  a long exploratory numerical study. This is beyond the scope of this $\it Letter$ and will be studied in subsequent papers. For that reason, we decided instead to adopt an idealize   model where star formation is not included and the thermodynamics is simplified by using an adiabatic Equation of State (E.O.S.), but where the velocity dispersion of the bulge  $\rm \sigma$ is the only varying parameter. With this E.O.S. we will not be able to compute accurate mass accretion rates, but we will be able to  compare the changes in the accretion rates for different values of $\rm \sigma$ and then reveal the role of  $\sigma$ in the mass transport due to gravity-driven turbulence.

The model consist in a $\rm M_{g} = 5 \times 10^{7} \msun$ gaseous disk  around a MBH $\rm (M_{BH} = 10^{7} \msun)$ that is embedded in a stellar bulge. The MBH and the bulge are modeled as  time-independent external potentials. For the bulge we use the potential of an isothermal sphere for $\rm r \geq 12 pc$, and the potential of an homogenous sphere for $\rm r < 12 pc$. We modeled the MBH by a Plummer potential with core radius of 1 pc. We solved the hydrodynamic and Poisson equations  using the FLASH adaptative mesh refinement hydrodynamics code (Fryxell et. al. 2000). The cartesian grid covers a  $\rm 256^{2} \times 32 \, pc^{3}$  region around the galactic center, with a  spatial resolution of 2pc. We use outflow boundary conditions.


The initial condition is a rotationally supported disk with uniform density profile, radius of 125 pc  and thickness of 4 pc. The initial temperature is set to $\rm 10^{3}$K over the whole region. Random density and temperature fluctuations of less than 1\% are added to the initial uniform disk. We performed four different runs where we vary the mass of the bulge, in  such a way that the velocity dispersion of the bulge  $\rm \sigma$ is: 100, 200, 300, and 400 $\rm km \, s^{-1}$. For a bulge modeled by an isothermal sphere, our rotationally supported disks have flat rotation curves. For example  the run with the highest gas fraction $\rm M_{gas}/M_{star}$  has a rotation velocity $\rm  v_{rot}$ of  109 $\rm km \ s^{-1}$. This corresponds to the $\rm \sigma =  100 km \, s^{-1}$ run and the $\rm M_{gas}/M_{star}$ ratio is similar to those found in the most massive gaseous nuclear disks (Downes \& Solomon 1998). 





\subsection{Results}

\begin{figure}
\plotone{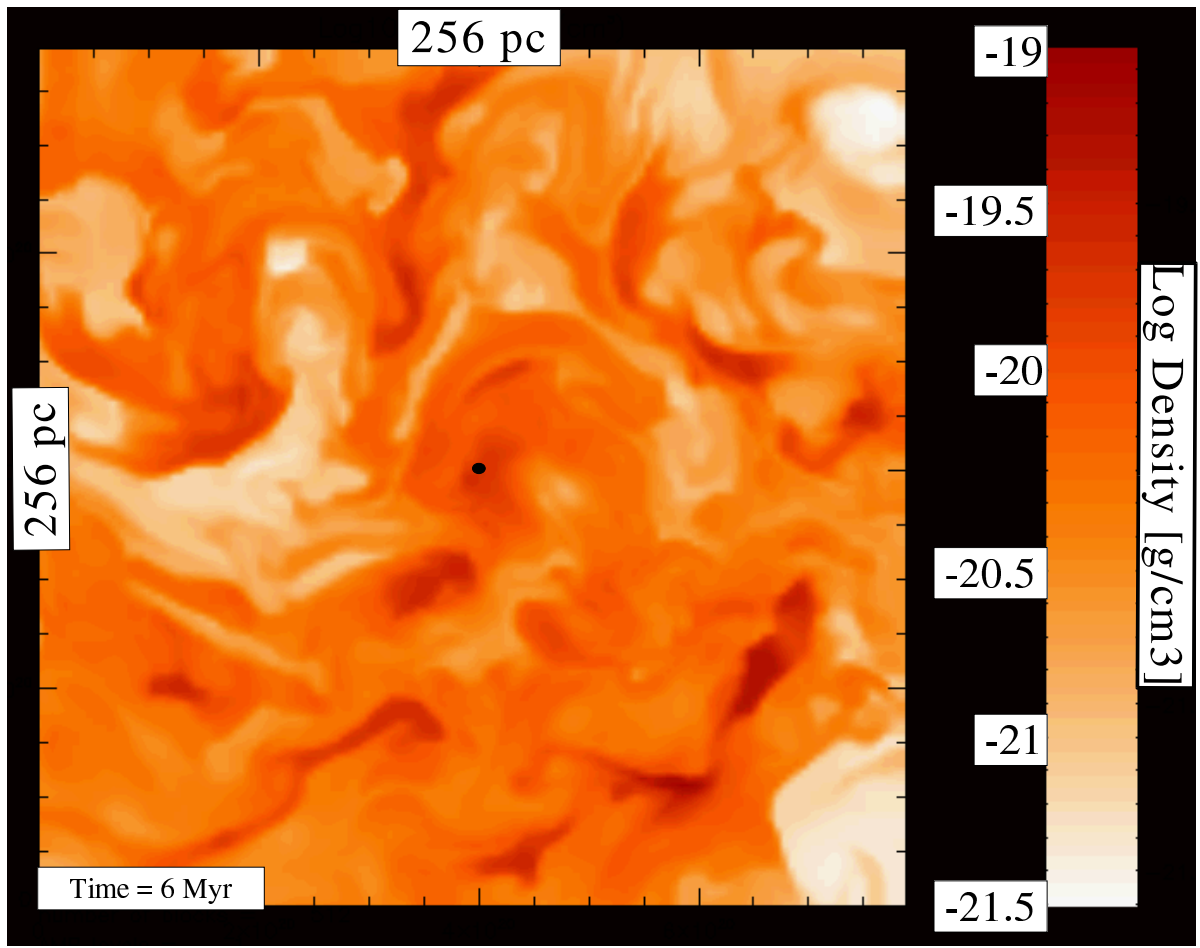}
\caption{Density distribution in the plane of the gas disk, coded on a logarithmic scale, at time t = 6Myr  for  the run 100. The black dot indicates the position of the MBH. The medium is characterized by high density clumps and filaments,  embedded in a less dense medium. }
\label{f1}
\end{figure}

Figure $\ref{f1}$ illustrates a representative stage in the evolution of the system, showing the face-on density distribution at the plane of the disk for  run 100, at the time t = 6 Myr. The figure shows a complicated multiphase  and  highly turbulent structure,   in qualitative  agreement with  the findings of  Wada \& Norman (2002). 

As  mentioned before, we start our simulation with an uniform disk with random  fluctuations of less than 1\% which are  non-linearly amplified by self-gravity. The system is then let to evolve   approximately for an orbital time $\rm t_{orb} = 2R_{d}/v_{rot}$, in order to fully develop the turbulence in the disk. Because  $\rm t_{orb}$ varies among the different runs, we choose the longer $\rm t_{orb}$ as starting point for our comparison. This corresponds to the run with  $\rm \sigma =  100 km \, s^{-1}$ which has an orbital time $\rm t_{orb} = 2.44 Myr$. 

\begin{figure}
\plotone{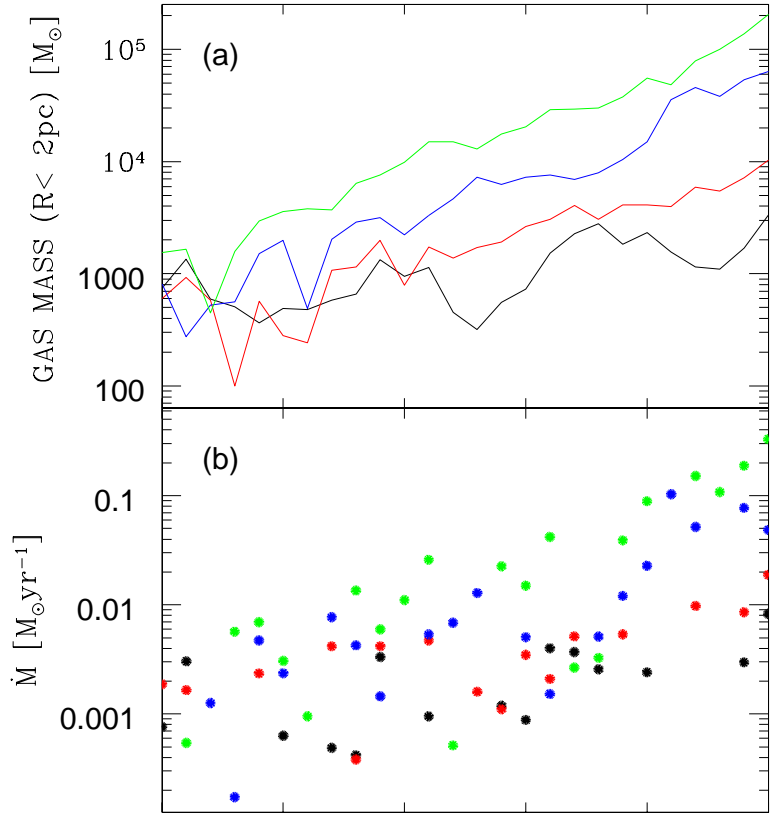}
\caption{(a) Time  evolution  of the gas mass inside 2 pc around the MBH (our resolution limit), for the four  different values of $\sigma$: 100 (black), 200 (red), 300 (blue) and 400  $\rm km \, s^{-1}$ (green). (b) Time evolution of the mass accretion rate inside 2 pc, $\rm \dot{M}$(r$<$2pc), using  the same color code as in (a).}

\label{f2}
\end{figure}

Figure $\ref{f2}$a shows the time  evolution  of the gas mass inside a radius of 2 pc around the MBH (our resolution limit), for the four  different values of $\sigma$: 100 (black), 200 (red), 300 (blue) and 400  $\rm km \, s^{-1}$ (green). This figure shows a strong dependence on the bulge velocity dispersion, and the final gas mass (r $<$ 2pc) varies by almost a factor 70 among the runs. In all the runs, the gas mass (r $<$ 2pc) shows an approximately  exponential growth. 

In Fig. $\ref{f2}$b, we plot the time evolution of the mass accretion rate inside 2 pc, $\rm \dot{M}$(r$<$2pc),  for the different runs and using the same color code as in  Fig. $\ref{f2}$a. The accretion rates are highly variable on timescales of $10^{5}$yr and with fluctuations of  a couple of orders of magnitude in $\rm \dot{M}$,  as  found by Wada (2004). However, we found accretion rates considerably lower than those of Wada (2004). The origin of this discrepancy rely in the adiabatic E.O.S. This assumption produces unrealistically high  Toomre's $\rm Q$  parameters compared to the observed ones ($\rm Q \sim 1.5$; Martin \& Kennicutt 2001),   decreasing the mass accretion rates on Eq. $\ref{mdot2}$.

\begin{figure}
\plotone{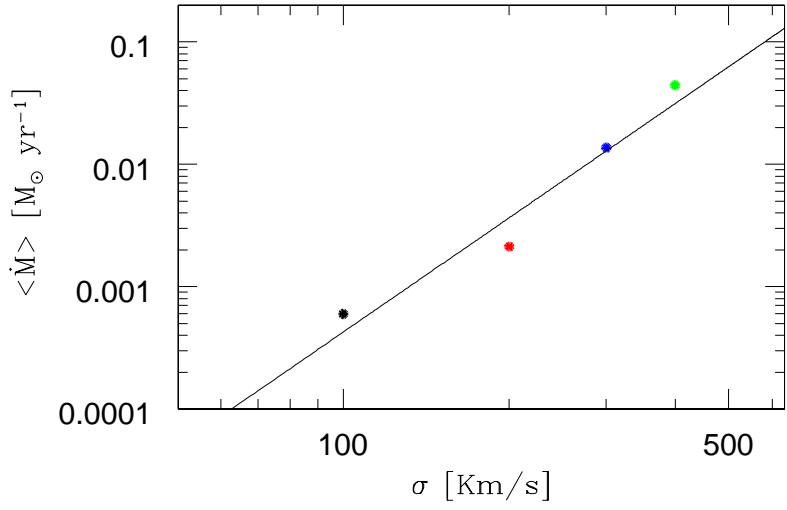}
\caption{Mass accretion rate averaged over the time interval [2.4Myr,7Myr], for the four  different values of $\sigma$: 100, 200, 300 and 400  $\rm km \, s^{-1}$ (in the same color code as in Fig. 2). The black line is the  best-fitting of the points using minimum least squares, and corresponds to  $\rm < \dot{M}> \propto \sigma^{3.1}$.} 
\label{f3}
\end{figure}

Finally, in Fig. $\ref{f3}$ we plot the average mass accretion rate $\rm <\dot{M}>$ as a function of $\rm \sigma$.  We compute  $\rm <\dot{M}>$ simply as $\rm [M(t_{f}) - M(t_{i})]/[t_{f}-t_{i}]$, where  $\rm t_{i}$ and  $\rm t_{f}$ are the initial and final times. We choose $\rm t_{i}$ = 2.4 Myr to ensure that  all the runs  have evolved for at least an orbital time $\rm t_{orb}$ and  at that time all the runs have almost the same   gas mass inside 2 pc. The final time of the runs is $\rm t_{f}$ = 7 Myr. It is found that the average mass accretion rate $\rm <\dot{M}>$ is strongly dependent on  $\rm \sigma$, as seen in  Fig. $\ref{f3}$.  The black line is a least squares fit to the points  in  Fig. $\ref{f3}$  and corresponds to  $\rm < \dot{M}> \propto \sigma^{3.1}$. Therefore  $\rm <\dot{M}>$ scales with $\rm \sigma$  as expected from Eq. $\ref{mdot2}$, despite the complex behavior of the gas (Fig. $\ref{f1}$) and that the mass accretion rates are highly time-dependent (Fig. $\ref{f2}$). 

\section{Discussion}

We perform a simple numerical experiment that illustrates the role of the spheroid  gravitational potential on the mass transport in a gravity-driven turbulent disk. We find that -on average- the mass accretion rate behaves as predicted by the zero-order approximation $\rm (\dot{M} \propto \sigma^{3})$, however our adiabatic simulations underestimate the mass accretion rates. 

Using more realistic models by   Wada \& Norman (2002) that found  $\rm \dot{M} \sim 1.4 \msun yr^{-1}$ for an spheroid with  $\rm \sigma = 100 \, kms^{-1}$, we estimate  that $\rm \dot{M} \approx  11.2 \, (\sigma/200 \, kms^{-1})^{3} \,  \msun yr^{-1}$ (Eq. $\ref{mdot2}$). If we assume  that 10\% of the mass fueled into the inner accretion disk is accreted by the MBH ($\rm \eta = 0.1$) and use the `$\rm M_{BH}$ - $\sigma$' relation, we find that MBHs could gained a mass $\rm M_{BH}$  in a rapid growth phase of $\rm t_{growth} \equiv M_{BH}/\eta\dot{M} \approx  10^{8} \, (M_{BH}/10^{8} \, \msun)^{1/4} \, yr$. This is in  agreement with  recent observations  (Borys et al. 2005), that are  consistent with a model where MBHs in submillimeter galaxies undergo a rapid growth to reach the local  `$\rm M_{BH}$ - $\sigma$' relation.

We find that the `$\rm M_{BH}$ - $\sigma$' relation arises naturally from the fueling of either  a turbulent disk or  a `disk of clouds' (coupled with the  Kennicutt-Schmidt Law), and therefore is hard to distinguish between the two models. However, the zero point could be  different allowing to discriminate between both disks.  Also, the zero point   allow us to address the possible role of star formation driven turbulence. Detailed numerical simulations of these various disks are necessary to determine a realistic zero point that could be compared with the one measured in the  `$\rm M_{BH}$ - $\sigma$' relation.


\bigskip

I thank Paolo Coppi for early stimulating discussions and   access to Yale's HPC facilities. I  thank Richard Larson, Diego Mardones and Paulina Lira for valuable comments. This research was funded by  FONDAP grant 15010003. The software used in this work was in part developed by the DOE-supported ASC / Alliance Center for Astrophysical Thermonuclear Flashes at the University of Chicago.

\newpage

\end{document}